\documentclass[%
a4paper,       
UKenglish,     
DIV10,          
useAMS,usenatbib,referee, doublespace]%
{scrartcl}

\usepackage[utf8]{inputenc}
\usepackage[UKenglish]{babel}
\usepackage[T1]{fontenc}
\usepackage{textcomp}
\usepackage{setspace}
\usepackage{hyperref}  
\usepackage{url}                
\usepackage[sumlimits, intlimits, namelimits]{amsmath} 
\usepackage{amssymb}            
\usepackage[caption = true]{subfig} 
\usepackage{array}              
\usepackage{booktabs}           
\usepackage{ae}                 
\usepackage[longnamesfirst,round]{natbib} 
\usepackage{color}              
\usepackage[sc]{mathpazo}       
\usepackage{helvet}             
\usepackage{todonotes}          

\def \blinded {1}  

\hypersetup{
  bookmarksopen=true, 
  breaklinks=true,
  pdftitle={The Assessment of Intrinsic Credibility},
  pdfsubject={Statistics},
  pdfkeywords={Statistics, Evidence},
  colorlinks=true,
  linkcolor=blue,
  anchorcolor=black,
  citecolor=teal,
  urlcolor=purple
}

\usepackage{xifthen}            
\usepackage{array}
\usepackage{amssymb}

\newcommand{\hl}{\textcolor{black}}

\newcommand{\blanco}[1]{  } 
\newcommand{\prep}{p_{\scriptsize \mbox{rep}}}

\newcommand{\latin}[1]{\textit{#1}}

\newcommand{\abk}[1]{\mbox{#1}\xdot}
\DeclareRobustCommand\xdot{\futurelet\token\Xdot}
\def\Xdot{%
  \ifx\token\bgroup.%
  \else\ifx\token\egroup.%
  \else\ifx\token\/.%
  \else\ifx\token\ .%
  \else\ifx\token!.%
  \else\ifx\token,.%
  \else\ifx\token:.%
  \else\ifx\token;.%
  \else\ifx\token?.%
  \else\ifx\token/.%
  \else\ifx\token'.%
  \else\ifx\token).%
  \else\ifx\token-.%
  \else\ifx\token+.%
  \else\ifx\token~.%
  \else\ifx\token.%
  \else.\ %
  \fi\fi\fi\fi\fi\fi\fi\fi\fi\fi\fi\fi\fi\fi\fi\fi%
}

\newcommand{\eg}{\abk{\latin{e.\,g}}}
\newcommand{\ie}{\abk{\latin{i.\,e}}}

\newlength{\halbebreite}
\DeclareMathOperator{\Nor}{N} 
  
\newcommand{\partialv}[3][1]{%
\ifthenelse{#1 = 1}{\frac{\partial #2}{\partial #3}}{\frac{\partial^{#1} #2}{\partial #3^{#1}}}
} 

\newcommand{\partials}[3][1]{%
\ifthenelse{#1 = 1}{\frac{d #2}{d #3}}{\frac{d^{#1} #2}{d #3^{#1}}}
} 

\newcommand{\dseps}[2][1]{%
\ifthenelse{#1 = 1}{\frac{d}{d #2}}{\frac{d^{#1}}{d #2^{#1}}}
}

\newcommand{\dsepv}[2][1]{%
\ifthenelse{#1 = 1}{\frac{\partial}{\partial #2}}{\frac{\partial^{#1}}{\partial #2^{#1}}}
}

\newcommand{\ml}[2][1]{
\ifthenelse{#1 = 1}%
 {\hat{#2}_{\scriptscriptstyle{\mathrm{ML}}}}%
 {\hat{#2}^{#1}_{\scriptscriptstyle{\mathrm{ML}}}}
}
\newcommand{\map}[2][0]{
\ifthenelse{#1 = 0}%
 {\hat{#2}_{\scriptscriptstyle{\mathrm{MAP}}}}%
 {\hat{#2}_{{\scriptscriptstyle{\mathrm{MAP}}_{#1}}}}
}
\newcommand{\mpm}[2][0]{
\ifthenelse{#1 = 0}%
 {\hat{#2}_{\scriptscriptstyle{\mathrm{MPM}}}}%
 {\hat{#2}_{{\scriptscriptstyle{\mathrm{MPM}}_{#1}}}}
}

\newcommand{\given}{\,\vert\,} 

\newcommand{\abs}[1]{\left\lvert#1\right\rvert} 

\IfFileExists{upquote.sty}{\usepackage{upquote}}{}

\usepackage{Sweave}
\begin{document}

\title{The Assessment of Intrinsic Credibility and a New Argument for $p<0.005$}
\ifcase\blinded 
\author{}
\or          
\author{Leonhard
  Held\\ 
  Epidemiology, Biostatistics
  and Prevention Institute (EBPI)\\ 
  and Center for Reproducible Science (CRS) \\
  University of Zurich\\ Hirschengraben 84,
  8001 Zurich, Switzerland\\ Email: \texttt{leonhard.held@uzh.ch}}
\fi

\maketitle

\begin{center}
\begin{minipage}{12cm}
\textbf{Abstract}: The concept of intrinsic credibility has been
recently introduced to check the credibility of ``out of the blue''
findings without any prior support.  A significant result is deemed
intrinsically credible if it is in conflict with a sceptical prior
derived from the very same data that would make the effect
non-significant.  In this paper I propose to use Bayesian
prior-predictive tail probabilities to assess intrinsic credibility.
For the standard 5\% significance level, this leads to a new $p$-value
threshold that is remarkably close to the recently proposed $p<0.005$
standard.  I also introduce the credibility ratio, the ratio of the
upper to the lower limit of a standard confidence interval for the
corresponding effect size.  I show that the credibility ratio has to
be smaller than 5.8 such that a significant finding is also
intrinsically credible.  Finally, a $p$-value for intrinsic
credibility is proposed that is a simple function of the ordinary
$p$-value and has a direct frequentist interpretation in terms of the
probability of replicating an effect. \\
\noindent
  \textbf{Key Words}: {Confidence Interval; Credibility Ratio; Intrinsic Credibility; 
    Prior-Data Conflict; \hl{$P$-value; Replication}; Significance Test}
\end{minipage}
\end{center}

\section{Introduction}

\hl{The so-called replication crisis of science has been discussed
  extensively within the scientific community \citep{Ioannidis2005,Begley2015}. One aspect of the 
problem is the widespread misunderstanding and misinterpretation of
basic statistical concepts, such as the $p$-value \citep{cohen:1994, Greenland2016}. This has lead to a major rethinking and
new proposals for statistical inference, such as
to lower the threshold for statistical significance
from the traditional 0.05 level to 0.005
\citep{Johnson2013,BenjaminEtAlinpress}.} The proposal
has created a lot of discussion in the scientific community and the
shortcut ``$p<0.005$'' has been even shortlisted and highly commended
in the Statistic of the Year competition by the Royal Statistical
Society, see \url{http://bit.ly/2yWFPbD}.

Two arguments for this step are provided in
\citet{BenjaminEtAlinpress}: The first is based on the Bayes factor,
the second is based on the false discovery rate. Both arguments are
actually not new, \cite{els:1963} have already emphasized that
the evidence of $p$-values around 0.05 against a point null
hypothesis, as quantified by the Bayes factor, is much smaller than
one would naively expect: ``Even the utmost generosity to the
alternative hypothesis cannot make the evidence in favor of it as
strong as classical significance levels might suggest.'' Likewise,
\cite{pmid526924} have already argued that the false positive
rate ``could be considerably reduced by increasing the sample sizes
and by restricting the allowance made for the $\alpha$ error,
which should be set to a 1\% level as a minimum requirement.''
\citet{BenjaminEtAlinpress} therefore propose to lower the threshold
for statistical significance to 0.005 and to declare results with
$0.05>p>0.005$ as ``suggestive'', emphasizing the need for replication.

In this paper I provide a new argument for this categorization
  into three levels of evidence. The approach is based on the concept
  of intrinsic credibility \citep{matthews:2017}, a specific
  reverse-Bayes method to assess the credibility of claims of new
  discoveries.  I review and refine the approach and show that, if you
  dichotomize p-values into ``significant'' and ``non-significant'',
  the proposed method naturally leads to a more stringent threshold
  for intrinsic credibility. For the standard 5\% significance level,
  the new \hl{$p$-value} thres\-hold is 0.0056, remarkably close to
  the recently proposed $p<0.005$ \hl{standard}.  
  
  To assess intrinsic credibility based on a confidence interval
  rather than a $p$-value, I propose the credibility ratio, the ratio
  of the upper to the lower limit of a standard confidence interval
  \hl{for the corresponding effect size}.  
  I show that the credibility
  ratio has to be smaller than 5.8 {to ensure that a significant
    finding is also intrinsically credible.}  In Section
  \ref{sec:sec1} I provide a brief summary of the Analysis of
  Credibility and the specific concept of intrinsic credibility. The
  latter is central for the derivation of a threshold for intrinsic
  credibility, as outlined in Section \ref{sec:sec2}.  

\hl{ Lowering the threshold of statistical significance is only a
  temporary measure to the replication crisis \citep{Ioannidis2018}. A
  more radical step would be to abandon significance thresholds
  altogether \citep{McShane2018}, leaving $p$-values as a purely
  quantitative measure of the evidence against a point null
  hypothesis. In this spirit I extend the concept of intrinsic
  credibility and propose in Section \ref{sec:sec3} the $p$-value for
  intrinsic credibility, $p_{IC}$.  This new measure can be used to
  quantify the evidence for intrinsic credibility -- without any need
  for thresholding -- and has a direct and useful interpretation in
  terms of the probability of replicating an effect
  \citep{Killeen:2005}. Intrinsic credibility is thus directly linked
  to replication, a topic of central importance in the current debate on research
  reproducibility \citep{Goodman2016}.  I close with some discussion
  in Section \ref{sec:sec4}.  }

\section{Analysis of Credibility}
\label{sec:sec1}

Reverse-Bayes approaches allow the extraction of the properties of the
prior distribution needed to achieve a certain posterior statement for
the data at hand.  The idea to use Bayes's theorem in reverse
originates in the work by IJ Good \citep{good:1950,good:1983} and is
increasingly used to assess the plausibility of scientific claims and
findings
\citep{greenland:2006,greenland:2011,held2013,Colquhoun:2017}.
\citet{matthews:2001,matthews:2001b} has proposed the Analysis of
Credibility, a specific reverse-Bayes method \hl{to challenge claims of ``significance'', 
  see \citet{matthews:2017} for more recent developments. }

Analysis of Credibility is based on a conventional confidence interval
of level \hl{$\gamma$}, say, for an unknown effect size $\theta$ with
lower limit $L$ and upper limit $U$, say.  In the following I assume
that both $L$ and $U$ are symmetric around the effect estimate $\hat
\theta$ (assumed to be normally distributed) and that both are either
positive or negative, \ie the effect is significant at significance
level $\alpha$ $\hl{=1-\gamma}$.  \citet{matthews:2001,matthews:2001b} proposed
assessing the credibility of a statistically significant finding by
computing from the data a sceptical prior distribution for the effect
size $\theta$, normal with mean zero, that - combined with the
information given in the confidence interval for $\theta$ - results in
a posterior distribution which is just non-significant at level
$\alpha$, \ie either the $\alpha/2$ or the $1-\alpha/2$ posterior
quantile is zero.  It can be shown that the limits $\pm S$ of the
corresponding equi-tailed prior credible interval at level
\hl{$\gamma$} are given by
\begin{equation}\label{eq:S}
S = \frac{(U-L)^2}{4 \sqrt{UL}},
\end{equation}
where $S$ is called the {\em scepticism limit} and the interval
$[-S,S]$ is called the {\em critical prior interval}.  Note that
\eqref{eq:S} holds for any level $\gamma$, not just for
the traditional 95\% level.

It is convenient to express 
the variance $\tau^2$ of the sceptical prior 
as a function of the variance $\sigma^2$ (the squared
standard error, assumed to be known) of the estimate $\hat \theta$,
the corresponding test statistic $t=\hat \theta/\sigma$, and
$z_{\alpha/2}$, the $1-\alpha/2$ quantile of the standard normal
distribution:
\begin{equation}\label{eq:tau2}
\tau^2 = \frac{\sigma^2}{t^2/z_{\alpha/2}^2 - 1},
\end{equation}
where $t^2>z_{\alpha/2}^2$ is required for significance at level
$\alpha$.  Equation \eqref{eq:tau2} shows that the prior variance
$\tau^2$ can be both smaller or larger than $\sigma^2$, depending on
the value of $t^2$. If $t^2$ is substantially larger than
$z_{\alpha/2}^2$, then the sceptical prior variance will be relatively
small, \ie a relatively tight prior is needed to make the significant result non-significant. 
If $t^2$ is close to $z_{\alpha/2}^2$ (\ie the effect is
``borderline significant''), then the sceptical prior variance will be
relatively large, \ie a relatively vague prior is sufficient to make
the significant result non-significant.

\begin{figure}
  \begin{center}
\setkeys{Gin}{width=0.575\textwidth}
\includegraphics{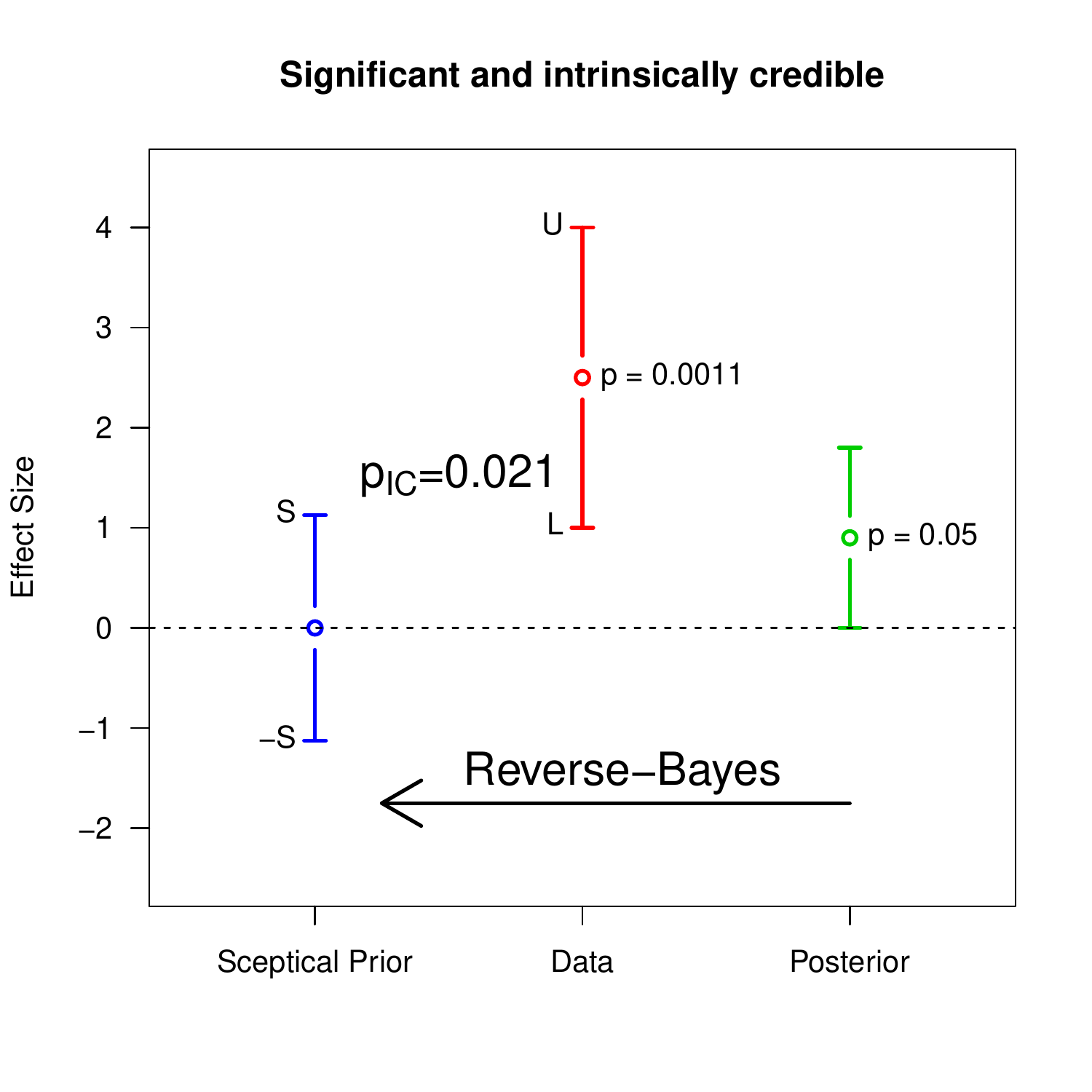}
\includegraphics{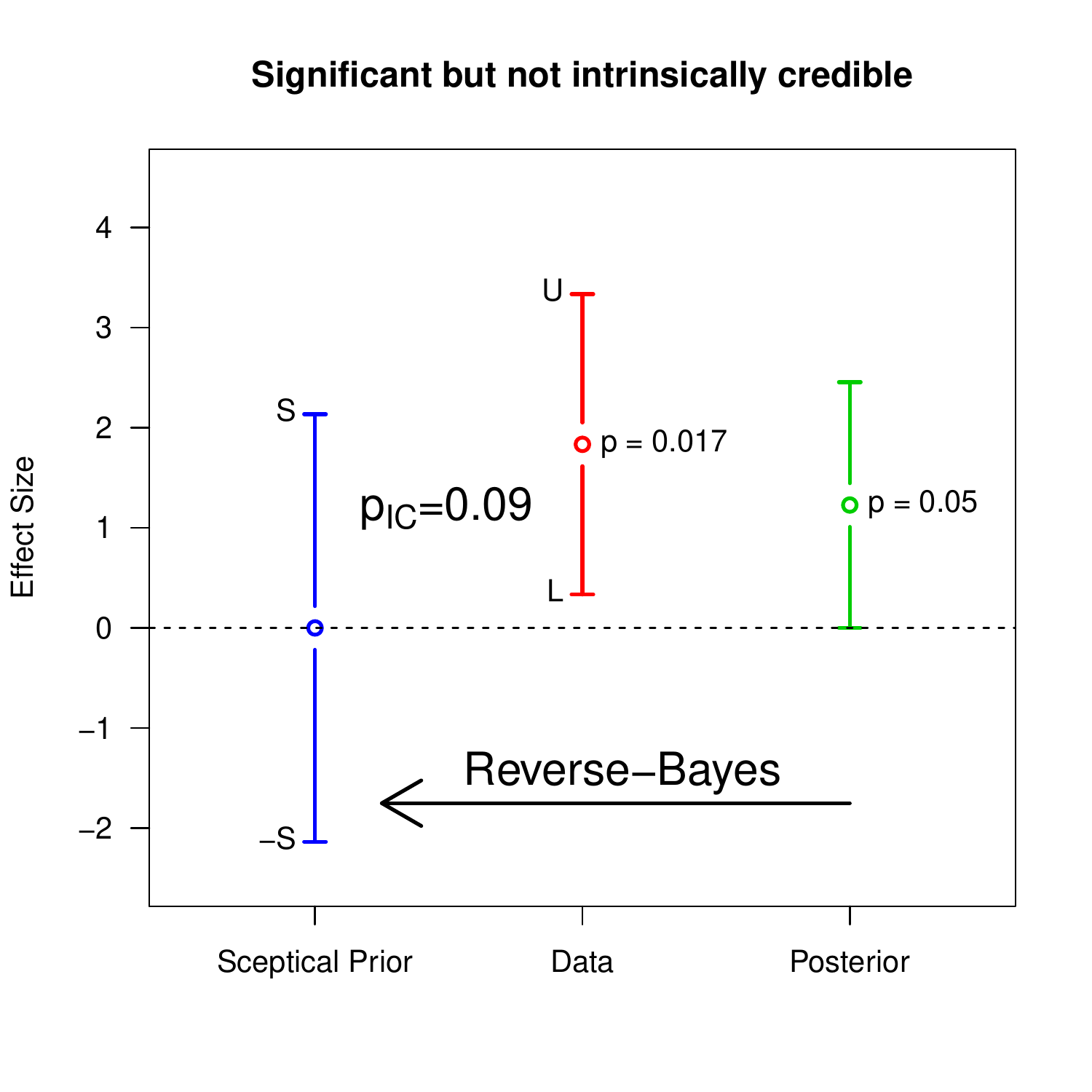}

\end{center}
\caption{Analysis of intrinsic credibility for two confidence intervals at level
  $\gamma=95\%$.  
    In the first example there is conflict between the sceptical prior
  and the data 
  and the significant result is
  \hl{intrinsically} credible \hl{at the 95\% level} ($L=1$, $U=4$, credibility ratio = 4, \hl{$p_{IC}=0.021$}).  In the second
  example there is \hl{less} conflict \hl{between prior and data} 
  and the significant result is not
  \hl{intrinsically} credible \hl{at the 95\% level}  ($L=1/3$, $U=10/3$, credibility ratio = 10, \hl{$p_{IC}=0.09$}). 
The credibility ratio
will be described further in Section \ref{sec:sec2} while the $p$-value $p_{IC}$ for intrinsic
credibility will be introduced in Section \ref{sec:sec3}.
    \label{fig:fig1}}
  
\end{figure}

\hl{Two applications of the Analysis of Credibility are shown in
  Figure \ref{fig:fig1}. Both are based on a confidence interval of
  width 3, but with different location ($\hat \theta = 2.5$ and $11/6=1.83$,
  respectively).
  Each Figure has to be read from right to left: To
  obtain a 95\% posterior credible interval with lower limit 0 (shown in green), the
  95\% confidence interval for the unknown effect size $\theta$ (shown in red) has to be combined with a sceptical
  prior with variance \eqref{eq:tau2} (shown in blue).}


\hl{ In this paper I focus on claims of new discoveries without any prior support. 
To assess the credibility of such ``out of the blue''  findings, 
  \citet{matthews:2017} suggested the concept of intrinsic
  credibility, declaring an effect as intrinsically credible if it is in conflict with the sceptical
  prior (with mean zero and variance \eqref{eq:tau2})
  that would make the effect
  non-significant.  This can be thought of as an additional check to ensure
  that a significant effect is not spurious.  Specifically,
  \citet{matthews:2017} declares a result as intrinsically credible at
  level $\gamma$, if the effect estimate $\hat \theta$ is outside the
  sceptical prior interval, \ie $\abs{\hat \theta} > S$.  }
He shows
that, for \hl{confidence intervals at level $\gamma=0.95$}, this is
equivalent to the conventional two-sided $p$-value being smaller than
0.0127.  \hl{I refine the definition of intrinsic credibility in the following Section
  \ref{sec:sec2} based on the \cite{box:1980} prior-predictive approach, 
    leading to the more stringent $p$-value threshold 0.0056 for intrinsic credibility at the 95\% level.}

\section{A new threshold for intrinsic credibility}
\label{sec:sec2}

Matthews' check for intrinsic credibility compares the size of $\hat
\theta$ with the scepticism limit \eqref{eq:S}, so does not take the uncertainty
of $\hat \theta$ into account. He compares the \hl{estimate $\hat \theta$}  with the
(sceptical) prior distribution, not with the \hl{corresponding} prior-predictive
distribution.  \hl{However, use of the latter is the established way to check the compatibility of the data and the prior \citep{box:1980,greenland:2006}}.
In
what follows I will 
\hl{therefore apply} the 
approach by \cite{box:1980} for the assessment of prior-data conflict \hl{based on the prior-predictive distribution},
with the perhaps slightly unusual feature that the prior has been
derived from the data. I argue that there is nothing intrinsically
inconsistent in investigating the compatibility of a prior, defined
through the data, and the data itself, extending an argument by
\citet[Section 5.10]{cox:2006} to the reverse-Bayes setting.

The \citet{box:1980} check for prior-data conflict
is based on the prior-predictive distribution, which is in our case
normal with mean zero and variance $\tau^2 + \sigma^2$ \citep[Section
  5.8]{sam:2004}.  The procedure
is based on the test statistic
$t_{\mbox{\scriptsize Box}} = {\hat \theta}/{\sqrt{\tau^2+\sigma^2}}$
and the (two-sided) tail probability 
$  p_{\mbox{\scriptsize Box}} =\Pr(\chi^2(1) \geq t_{\mbox{\scriptsize Box}}^2)$
as the corresponding upper tail of a
$\chi^2$-distribution with one degree of freedom. 
Small values of $p_{\mbox{\scriptsize Box}}$ indicate a conflict between the sceptical prior
and the data.

\hl{Now suppose we fix the confidence level at the conventional 95\% level, \ie $\gamma=0.95$.}
 Intrinsic credibility at the 95\% level (\ie $p_{\mbox{\scriptsize Box}} < 0.05$) can
then be shown to be equivalent to the requirement $p<0.0056$ for the conventional two-sided
$p$-value.  
To derive this result, note that with \eqref{eq:tau2} we have
$\tau^2 + \sigma^2 = \sigma^2 / (1- z_{\alpha/2}^2 / t^2 )$
and so 
 $t_{\mbox{\scriptsize Box}}^2  
  =  t^2 - z_{\alpha/2}^2$. 
The requirement $t_{\mbox{\scriptsize Box}}^2 > z_{\alpha/2}^2$ for intrinsic
credibility at level $\gamma=1-\alpha$ then translates to
\begin{eqnarray}\label{eq:int.cred}
t^2 & \geq &  2 \, z_{\alpha/2}^2. 
\end{eqnarray}
This criterion is to be compared with the traditional check for
significance, which requires only $t^2 \geq z_{\alpha/2}^2$.
\hl{It follows directly that
the threshold}
\begin{eqnarray}
\alpha_{IC} &=& 2 \left\{1 - \Phi\left(t=\sqrt{2} \, z_{\alpha/2} \right) \right\}, \label{eq:iP}
\end{eqnarray}
\hl{here $\Phi(.)$ denotes the cumulative standard normal distribution
  function, } \hl{can be used to assess intrinsic credibility based on
  the conventional two-sided $p$-value $p$:} If $p$ is smaller than
$\alpha_{IC}$, then the result is intrinsically credible at level
$\gamma=1-\alpha$.  For $\alpha=0.05$ we have $t=\sqrt{2}\cdot 1.96=
2.77$ and the threshold \eqref{eq:iP} turns
out to be $\alpha_{IC} = 0.0056$, as claimed
above. \hl{For other confidence levels we will obtain other intrinsic
  credibility thresholds. For example, \citet[Section 10.1]{claytonhills} prefer to
  use 90\% confidence intervals ``on the grounds that they give a better
  impression of the range of plausible values''. Then $\gamma=0.9$ and
  we obtain the intrinsic credibility threshold $\alpha_{IC} =
  0.02$.}

\begin{center}
\begin{figure}[ht]
\begin{center}
\includegraphics{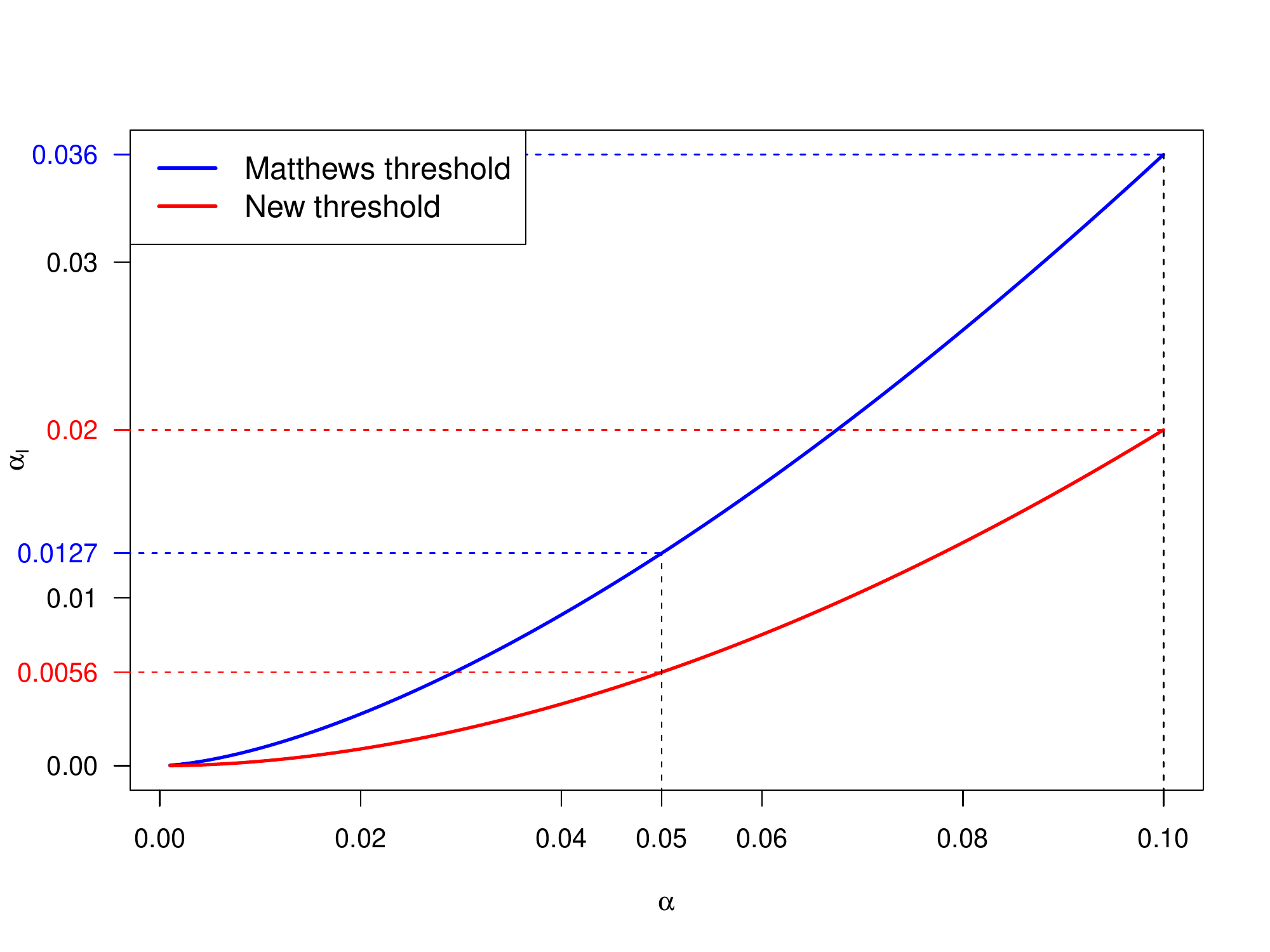}
\caption{The threshold for intrinsic credibility of significant results as a
  function of the conventional $\alpha$ level. The blue line
  corresponds to the proposal by \citet{matthews:2017}. The red line is the proposed new threshold.} 
\label{fig:fig2}
\end{center}
\end{figure}
\end{center}

Figure \ref{fig:fig2} compares the new threshold with the one obtained
by \citet[Appendix A.4]{matthews:2017} (using
$t=1.272 \, z_{\alpha/2}$) for values of
$\alpha$ below 10\%.  
The Matthews threshold for intrinsic credibility
is larger than the proposed new threshold \eqref{eq:iP}, because it compares 
\hl{the effect estimate $\hat \theta$} with the prior distribution \hl{(with variance $\tau^2$)} 
and 
not the prior-predictive distribution \hl{(with variance $\tau^2 + \sigma^2$)}.

\hl{Intrinsic credibility can also be assessed based on the
  confidence interval $[L, U]$, rather than the conventional $p$-value $p$.}
{\hl{To see this, note that }$t_{\mbox{\scriptsize Box}}^2$
can be written in terms of $L$ and $U$, 
\begin{equation}\label{eq:tBoxNew}
t_{\mbox{\scriptsize Box}}^2 = z_{\alpha/2}^2 \frac{4 \, U L}{(U-L)^2},
\end{equation}
and the requirement $t_{\mbox{\scriptsize Box}}^2 \geq z_{\alpha/2}^2$ for 
intrinsic credibility is then equivalent to require
that the {\em credibility ratio} $U/L$ (or $L/U$ if both $L$ and $U$ are negative) fulfills
\begin{equation}\label{eq:cr}
\frac{U}{L} \leq d = 3 + 2 \, \sqrt{2} \approx 5.8.
\end{equation}
To
derive the cut-point $d$ in \eqref{eq:cr}, set $U=L \, d$. The
requirement $t_{\mbox{\scriptsize Box}}^2 = z_{\alpha/2}^2$ then reduces to 
$$1 = \frac{4 \, U L}{(U-L)^2} = \frac{4 \, d}{(d-1)^2},$$ a quadratic
equation in $d$ with $d=3 + 2 \, \sqrt{2}$ 
as solution. 

Thus, there is a \hl{second} way to assess intrinsic credibility based
on the ratio of the limits of a confidence interval {\em at any level}
\hl{$\gamma$}: if the credibility ratio is smaller than
$5.8$ than the result is credible at level
\hl{$\gamma$}. 
\hl{For example, in Figure \ref{fig:fig1} the credibility
ratio is 4 in the top and 10 in the bottom panel, so the result
shown in the top panel is intrinsically credible at level 95\%, but the one in the bottom is not.}

If the sceptical prior distribution is available, then \hl{a third}
way to assess intrinsic credibility is to compare the prior variance
$\tau^2$ to the data variance $\sigma^2$. Comparing \eqref{eq:tau2}
with \eqref{eq:int.cred} it is easy to see that intrinsic credibility 
is achieved if and only if the sceptical prior variance $\tau^2$ is not larger
than the variance $\sigma^2$ of the effect estimate $\hat \theta$. With
this in mind we see immediately from Figure \ref{fig:fig1} that the first
result shown in the top panel is intrinsically credible ($\tau^2 < \sigma^2$), whereas the
second isn't ($\tau^2 > \sigma^2$).

\section{\hl{A $p$-value for intrinsic credibility}}
\label{sec:sec3}

\hl{ A disadvantage of the dichotomous assessment of intrinsic
  credibility described in the previous section is the dependence on
  the confidence level \hl{$\gamma$} of the underlying confidence
  interval, or, equivalently, the significance level
  $\alpha=1-\gamma$.  However, there is a way to free ourselves from
  this dependence.  In analogy to the well-known duality of confidence
  intervals and standard $p$-values, I propose to derive the value
  $\alpha^\star$, say, that just achieves intrinsic credibility, \ie
  where equality holds in \eqref{eq:int.cred}.  This defines the {\em
    $p$-value for intrinsic credibility} $p_{IC}=\alpha^\star$, which
  provides a quantitative assessment of the evidence for intrinsic
  credibility. Of course, the $p$-value for intrinsic credibility
  $p_{IC}$ can also be used to assess intrinsic credibility as
  described in Section \ref{sec:sec2}: if $p_{IC} \leq \alpha$, then
  the result is intrinsically credible at level $\gamma=1-\alpha$.}

\hl{
The $p$-value $p_{IC}$ for intrinsic
credibility can be derived by replacing $\alpha_{IC}$
with $p$ and $\alpha$ with $p_{IC}$ in equation \eqref{eq:eq3} and then solving for $p_{IC}$:
\begin{equation}\label{eq:eq3}
p_{IC}=2\left[1-\Phi\left(t/\sqrt{2} \right)\right].
\end{equation}
Here $t= \Phi^{-1}(1-p/2)$ is the standard test statistic for
significance where $p$ is the conventional two-sided
$p$-value. Note that the test statistic $t_{I} = t /
\sqrt{2}$ for intrinsic credibility in \eqref{eq:eq3} is a root-2 shrunken version of the
test statistic $t$ for significance.
}

\hl{ Figure \ref{fig:fig3} shows that the $p$-value $p_{IC}$ for
  intrinsic credibility is considerably larger than the conventional $p$-value $p$, 
  particularly for small values of $p$.  For example, the two confidence intervals shown 
  in Figure  \ref{fig:fig1} have 
  conventional $p$-values $p=0.0011$ (top) and $p=0.017$ (bottom),  
  while the corresponding
  $p$-values for intrinsic credibility are $p_{IC}=0.021$
  and $p_{IC}=0.09$,
  respectively. If we are prepared to adapt the ``rough and ready'' $p$-value guide
  by \citet[Section 9.4]{Bland:2015} to $p_{IC}$, then
  $p_{IC}=0.021$ provides moderate
  evidence and $p_{IC}=0.09$ only weak
  evidence for intrinsic credibility. 
}

\begin{center}
\begin{figure}[ht]
  \begin{center}

\includegraphics{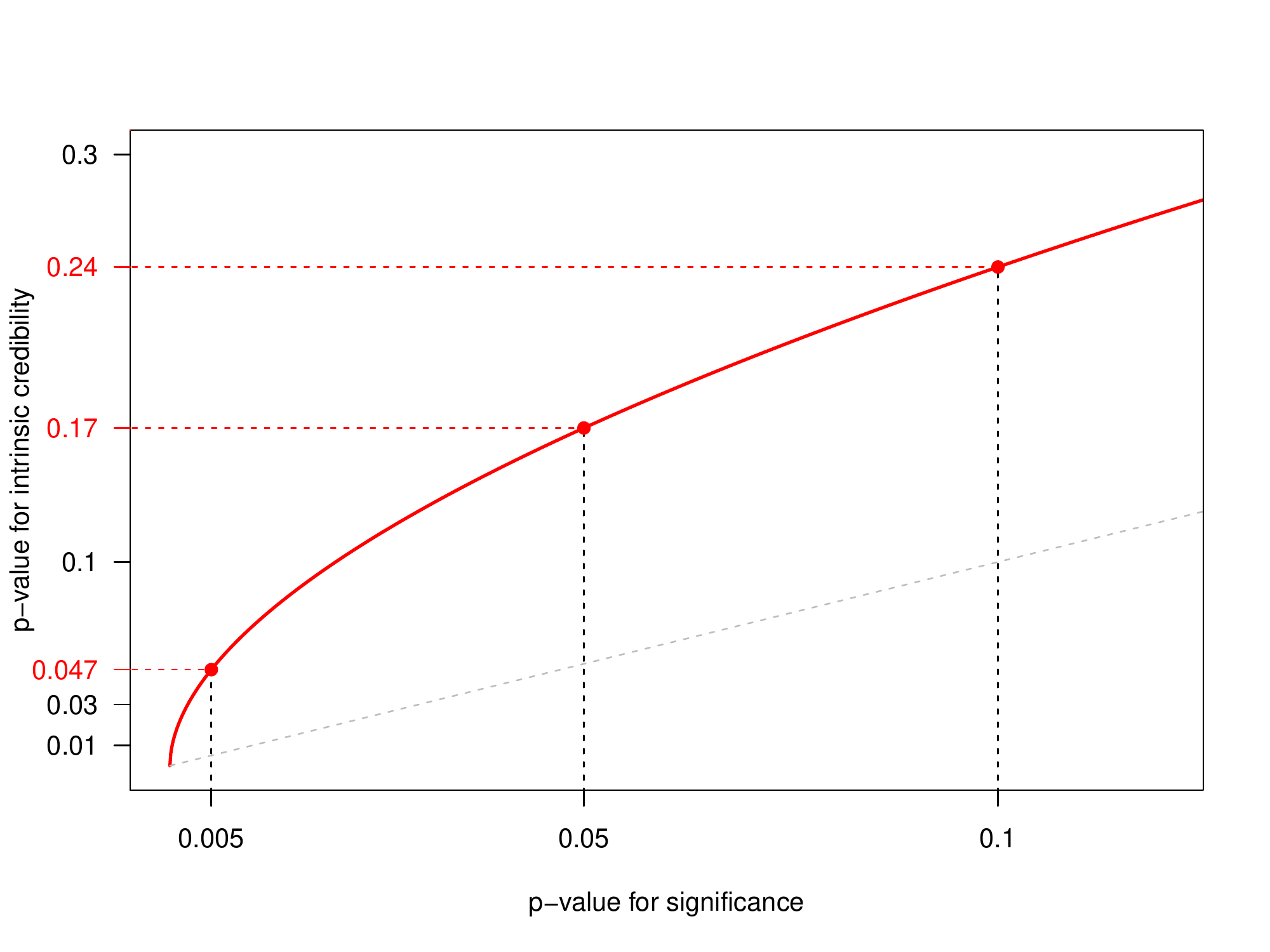}
\caption{The $p$-value for intrinsic
credibility as a function of the $p$-value for
significance. The grey dashed line is the identity line.}
\label{fig:fig3}
\end{center}
\end{figure}
\end{center}

  
  There is a direct and useful interpretation of $p_{IC}$  
  in terms of the probability of replicating an effect \citep{Killeen:2005}, \ie the probability 
  that an identically designed but independent replication study 
will give an estimated effect $\hat \theta_2$ in the
same direction as the estimate $\hat \theta_1 = \hat \theta$ from the current
(first) study. To see this, note that under an initial uniform prior
the posterior for $\theta$ is $\theta \given \hat \theta_1 \sim \Nor(\hat \theta_1,
\sigma^2)$. This posterior now serves as the prior for the mean of the (unobserved) estimate
$\hat \theta_2 \given \theta \sim \Nor(\theta, \sigma^2)$ from the second (hypothetical) study,
where we assumed the two studies to be identically designed, having equal
variances $\sigma^2$.  This leads to the prior-predictive distribution
$\hat \theta_2  \given \hat \theta_1 \sim \Nor(\hat \theta_1, 2 \, \sigma^2)$ and the $p$-value 
for intrinsic credibility \eqref{eq:eq3} can be seen to be
twice the probability that the second study will give an 
estimate $\hat \theta_2$ in the opposite direction as the estimate $\hat \theta_1$ of the 
first study:
\begin{eqnarray*}
p_{IC} &=& 2\, \left[1-\Phi\left({t}/{\sqrt{2}} \right) \right] \\
&=& 2\, \Phi\left({-t}/{\sqrt{2}} \right) \\
&=& 2\, \Phi\left(\frac{0 - \hat \theta_1}{\sqrt{2}\sigma} \right) \\
&=&  2\, \Pr(\hat \theta_2 \leq 0 \given \hat \theta_1 > 0).
\end{eqnarray*}
If $\hat \theta_1 < 0$, then $p_{IC} = 2\, \Pr(\hat \theta_2 \geq 0 \given \hat
\theta_1 < 0)$.

\hl{ 
  The probability $\Pr(\hat \theta_2 \leq 0 \given \hat \theta_1 >
  0)$ is one of the three replication probabilities that have been
  considered by \citet{SIM:SIM1072} in response to 
  \citet{SIM:SIM4780110705}. The complementary probability $\Pr(\hat
  \theta_2 > 0 \given \hat \theta_1 > 0) = 1 - p_{IC}/2$ can be
  identified as the probability of replicating an effect, $\prep$,  
  advocated by \cite{Killeen:2005}
  as an alternative to traditional $p$-values, see
\citet{LecoutrePoitevineau2014,Killeen2015} for further discussion and additional references. 
  Of course, $\prep$ can only be correct under the assumption that the null hypothesis is false. 
  Nevertheless,
  \cite{Killeen:2005,Killeen2015} argues that $\prep$ is a useful alternative to traditional $p$-values.
}

\hl{In practice, we can thus use $p_{IC}$ to assess the probability of
  replicating an effect, assuming that the null hypothesis is false: $\prep = 1 - p_{IC}/2$. An intrinsically
  credible result with $p_{IC} \leq \gamma$ therefore has $\prep \geq (1+\gamma)/2$. 
  For example, for $\gamma=95\%$ we have $\prep \geq 97.5\%$. 
For numerical illustration, recall that the $p$-values for
intrinsic credibility in Figure \ref{fig:fig1} are
$p_{IC}=0.021$ (top) and
$p_{IC}=0.09$
(bottom). The corresponding replication probabilities are thus
$\prep=99.0\%$ and
$\prep=95.5\%$.  In
the second example, there is thus a
$\prep=4.5$\%
chance that an identically designed replication study will give a negative effect estimate.
}

\section{Discussion}
\label{sec:sec4}

Based on the Analysis of Credibility, I have \hl{shown that, if you
  dichotomize p-values into ``significant'' and ``non-significant'' at
  some pre-specified threshold $\alpha$, the Analysis of Credibility
  directly leads to a more stringent threshold $\alpha_{IC}$ for
  intrinsic credibility. If you prefer to avoid any thresholding of conventional $p$-values,
  a new $p$-value for intrinsic credibility, $p_{IC}$, has been proposed.    
  $p_{IC}$ is a quantitative measure of the evidence for intrinsic credibility with a
  direct connection to $\prep$, the probability of replicating an effect
\citep{Killeen:2005}.}

The assessment of intrinsic credibility can
be thought of as an \hl{additional challenge,  
ensuring that claims of new
  discoveries} are not spurious. Conventionally significant results
with $0.05 > p > 0.0056$ 
lack intrinsic credibility, \ie they are not in conflict with a
sceptical prior that would make the effect non-significant. This
matches the classification as ``suggestive'' by
\cite{BenjaminEtAlinpress}. Specifically, $p > 0.0056$ implies 
$p_{IC} > 0.05$ and thus $\prep < 97.5\%$, 
emphasizing the need for replication.
If $p<0.0056$, then the result
is both significant and intrinsically credible at the 95\% level, 
\hl{so $p_{IC} \leq 0.05$ and $\prep \geq 97.5\%$}.

The credibility ratio provides a simple and convenient tool to check
whether a ``significant'' confidence interval at any level \hl{$\gamma$}
is also intrinsically credible. If
the credibility ratio is smaller 5.8, the result can be considered
as intrinsically credible at level \hl{$\gamma$}.
It is noteworthy that the \hl{concept of intrinsic credibility} does not require to change the
original confidence level $\gamma$. Indeed, the check for credibility is done
at the same level as the original confidence level. I have used
$\gamma=0.95$ by convention, where it follows that the check for
intrinsic credibility is equivalent to the requirement $p<0.0056$.
\hl{This implies that in standard statistical reporting there is no need to
replace 95\% confidence intervals with 99.5\% confidence intervals, say. 
However, I suggest to complement or to replace the ordinary $p$-value
with the proposed $p$-value for intrinsic credibility, $p_{IC}$.
}

Although derived using a Bayesian approach, the proposed check for
intrinsic credibility is based on a standard confidence interval and
thus constitutes a {Bayes/non-Bayes compromise} \citep{MR1185188}.
Specifically, it does not require the specification of a prior
probability of the null hypothesis of no effect. \hl{In fact, this
  prior probability is always zero.}  This is in contrast to the
calibration of $p$-values to lower bounds on the posterior probability
of the null, which requires specification of a prior probability.
Minimum Bayes factors have also been proposed to calibrate $p$-values,
see \citet{HeldOtt2018} for a recent review.  \hl{They have the
  advantage that they do not require specification of a prior
  probability of the null hypothesis and provide a direct
  ``forward-Bayes'' assessment of the evidence of $p$-values. However,
  the underlying rationale is still based on a point null hypothesis with
  positive prior probability, fundamentally different from the
  approach proposed here.  }


The Analysis of Credibility assumes a simple mathematical framework,
where likelihood, prior and posterior are all normally
distributed. \hl{This can be justified because Gaussian
  approximations are commonly used in the calculation of confidence
  intervals and statistical hypothesis tests, if the sample size is fairly large \citep[\eg][Section
    2.4]{sam:2004}. Of course, suitable transformations of the
  parameter of interest may be needed to achieve normality, for
  example, confidence intervals for odds ratios and hazard ratios should be transformed to the
  log scale. For small studies, however, the normal assumption for the likelihood may be questionable and the 
  assessment of intrinsic credibility would need appropriate refinement, for example based on the $t$-distribution.}

\ifcase\blinded 
{}
\or
{
\section*{Acknowledgments}
I am grateful to Robert Matthews, Stefanie Muff, Manuela Ott and Kelly Reeve 
for helpful comments on earlier drafts of this manuscript. 
} 
\fi

\bibliographystyle{apalike}
\bibliography{antritt,repro}
\end{document}